# Contents

# Formation of Giant Planets

Gennaro D'Angelo and Jack J. Lissauer

**Abstract** Giant planets are tens to thousands of times as massive as the Earth, and many times as large. Most of their volumes are occupied by hydrogen and helium, the primary constituents of the protostellar disks from which they formed. Significantly, the solar system giants are also highly enriched in heavier elements relative to the Sun, indicating that solid material participated in their assembly. Giant planets account for most of the mass of our planetary system and of those extrasolar planetary systems in which they are present. Therefore, giant planets are primary actors in determining the orbital architectures of planetary systems and, possibly, in affecting the composition of terrestrial planets. This Chapter describes the principal route that, according to current knowledge, can lead to the formation of giant planets, the core nucleated accretion model, and an alternative route, the disk instability model, which may lead to the formation of planetary-mass objects on wide orbits.
*A PDF version of this manuscript with working hyperlinks can be downloaded from* https://figshare.com/s/8ef727eb9f760582ae6c

**Key words:** Jupiter – Jovian planets – planetary formation – planet-disk interaction – disk self-gravity

## Introduction

Giant planets are so named because they are much larger and more massive than the Earth and the other solar system terrestrial planets. Jupiter, the prototype of giant

Gennaro D'Angelo
Los Alamos National Laboratory, Theoretical Division, MS B216, Los Alamos, NM 87545, USA, e-mail: gennaro@lanl.gov

Jack J. Lissauer
NASA Ames Research Center, Space Science & Astrobiology Divison, MS 245-3, Moffett Field, CA 94035, USA, e-mail: jack.lissauer@nasa.gov

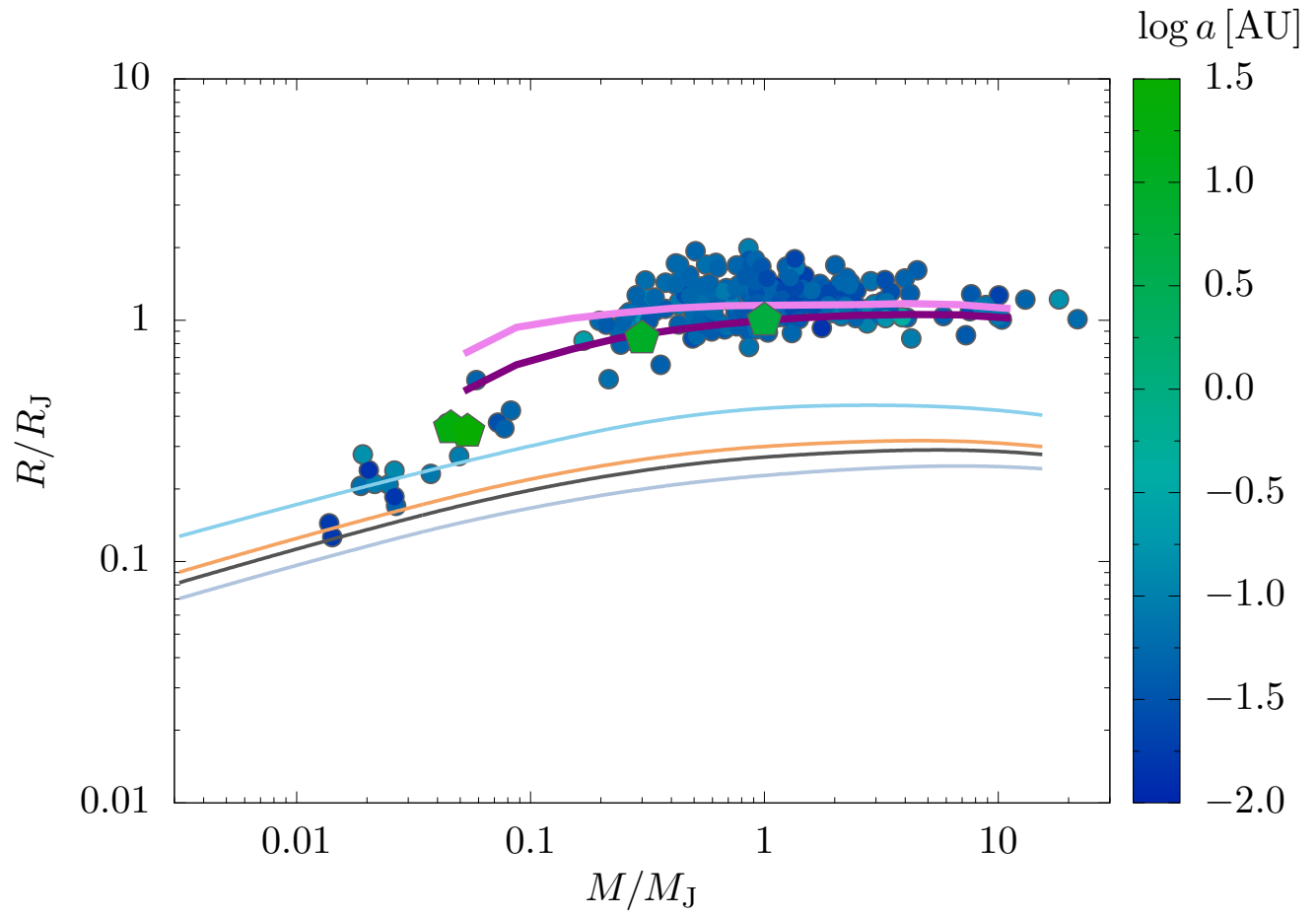

**Fig. 1** Radius versus mass of solar system (pentagons) and extrasolar (circles) planets ($M_J = 317.8\,M_\oplus$, $R_J = 10.97\,R_\oplus$). The logarithm of the orbital distance, $a$, is rendered by the color scale. The selection includes only well-characterized, confirmed extrasolar planets (source: http://exoplanets.org). The four thinner curves represent the computed radii of condensed planets (i.e., containing no H/He) of different compositions. From top to bottom, the composition is: 100% $H_2O$, Earth-like, Mercury-like, and 100% Fe (D'Angelo and Bodenheimer 2016). The two thicker curves represent the computed radii of 4.5 Gyr-old giant planets, orbiting a sun-like star at $a = 0.02$, AU (upper curve) and 9.5 AU (from Fortney et al. 2007).

planets, has a volumetric mean radius 10.97 times the Earth's radius ($R_\oplus$) and is 317.8 times the Earth's mass ($M_\oplus$). The radius of Saturn is $9.14\,R_\oplus$ and the mass is $95.2\,M_\oplus$. These planets must be gaseous, or else they would not be so large. Figure 1 compares observations, including the solar system gas and ice giants (Uranus and Neptune), to theoretical mass-radius curves of condensed planets of various compositions and of giant planets with hydrogen and helium in solar proportions (see the figure's caption for further details). As illustrated in the figure, only gas-dominated planets can achieve radii $R \gtrsim 7\,R_\oplus$. The radius of a "solid" planet increases as $R \propto M^{1/3}$ at low masses. However, beyond a certain mass, $R$ ceases to increase as $M$ increases because of the effects of electron degeneracy on pressure (e.g., Seager et al. 2007). The figure also shows that giant planets can be significantly larger than Jupiter, implying a hotter interior (e.g., Fortney and Nettelmann 2010; Baraffe et al. 2014).

Giant planets undergo a distinctive evolutionary phase, during which they accrete hydrogen and helium from the surrounding environment on a (relatively) short timescale. Although the mass of Jupiter does not represent a standard value, as can be seen in Figure 1, observations do show that planets several times Jupiter's mass are rarer than planets up to a few times the mass of Jupiter. This result is a likely outcome of their formation process and of the environment in which they form. By virtue of their large gravity field, giant planets can have a profound impact on the evolution, properties, and architecture of their planetary system, just like Jupiter and Saturn had on the solar system. For example, there is some evidence that the collisional histories of the terrestrial planets would have been different without the solar system giants on their current orbits.

The scope of this Chapter is to present the basic knowledge on the formation of giant planets. Some fundamental constraints, provided or implied by observations, that must be satisfied by any proposed formation scenarios are listed in the next section. The most successful formation mechanism, the *core nucleated accretion model*, and an alternative mechanism, the *disk instability model*, are outlined in the two following sections. Persisting challenges to these formation scenarios are discussed in the final section.

## Overview of the Observations

Giant planets contain large amounts of hydrogen and helium, in nearly stellar proportions. In the solar system, most of the mass of Jupiter (around 85%, Fortney and Nettelmann 2010) and Saturn (probably around 75%, Fouchet et al. 2009) is accounted for by these two elements. Consequently, the formation timescale of giant planets must be shorter than, or equal to, the lifetime of the *gaseous* protoplanetary disk in which they formed (although some heavy-element material – i.e., $Z > 2$ – can be collected and some light-element constituents can be lost afterward). Current observations indicate that, within several astronomical units of solar-type stars, the gas phase of protoplanetary disks lasts a few to several million years (Myr, e.g., Hillenbrand 2008; Roberge and Kamp 2010; Ercolano and Pascucci 2017), and possibly somewhat longer (Bell et al. 2013).

Estimates of the condensed core mass of Jupiter range from $\approx 0$ to $\approx 18\,M_\oplus$ (e.g., Saumon and Guillot 2004; Militzer et al. 2008; Nettelmann et al. 2012). Current analysis of available data from the Juno Mission (Bolton et al. 2017) appears compatible with values up to $25\,M_\oplus$ (Wahl et al. 2017). For Saturn, values range from $\approx 15$ to $\approx 20\,M_\oplus$ (Hubbard et al. 2009). Yet, during formation, the actual condensed core masses may have been different and the total amount of astrophysical metals contained in these planets may be more relevant instead. The mass fraction of some heavy elements in Jupiter is a few to several times solar, and atmospheric measurements indicate that Saturn is even more enriched in elements such as C, S, and P (Atreya et al. 2016). Furthermore, observations suggest that at least some extrasolar giant planets have super-stellar abundances of heavy elements as well (e.g., Sato et al. 2005; Miller and Fortney 2011; Jordán et al. 2014).

All solar system gas and ice giants appear to have formed beyond the water condensation front, whereas extrasolar giants seem ubiquitous in terms of distances from their stars (see Figure 1). Although it is not known whether all these planets formed at their current orbital locations, it is widely accepted that those orbiting well inside the water condensation front, where the gas temperature was higher than $\approx 200\,\mathrm{K}$, formed at greater distances and were relocated during or after formation (see the review Chapters by C. Mordasini and by R. Pudritz et al.). Distant giant planets too may have undergone some degree of orbital migration, imprinted in the orbital configurations of other bodies in the system. Moreover, while all giant planets in the solar system have low-eccentricity orbits perpendicular to the Sun's spin vector, this is not the case for all extrasolar giants, which can have very large orbital eccentricity and inclinations, likely acquired during or soon after formation.

As Jupiter and Saturn, extrasolar giants too can have companions, even though most of them appear to be singles: only around 20% of the planets whose mass is $\gtrsim 0.3\,M_J$ are currently known to reside in a multi-planet system (source: http://exoplanets.org), although the sample is still incomplete. Unseen companions may, and probably do, exist in a number of cases. And giant companions might have been more numerous in the past. In fact, the processes that generated Hot/Warm Jupiters or that excited orbital eccentricities to large values may as well have caused, through scattering or collisions, the loss of some or even most of the planets' former siblings.

## Formation by Core Nucleated Accretion

Giant planet formation via core nucleated accretion begins with the assembly of a planetary embryo, a condensed core of heavy-element materials, which in the classical model grows out of approximately 1–100 km size bodies, referred to as planetesimals. Although the processes by which dust grains carried by the gas in a protoplanetary disk coagulate into larger particles, eventually forming planetesimals, are not yet fully understood, asteroids, Kuiper-belt objects, and comets observed today belong to this ancient class of bodies.

Pairwise collisions among planetesimals lead to the growth of a planetary embryo, which accretes smaller bodies in the proximity of its orbit, eventually becoming a planetary core. When the gravitational energy at the surface becomes larger in magnitude than the thermal (plus the relative kinetic) energy of the nearby gas, a core can accrete an atmosphere. The accretion of a H/He envelope distinguishes gas-rich planets, such as the solar system ice giants Uranus and Neptune and most extrasolar planets larger than $\approx 1.6\,R_\oplus$ in radius, from terrestrial-type, condensed planets. The composition of gas-rich planets is still dominated by heavy elements, although they contain significant amounts of H and He by mass. Planets that become gas-rich relatively early during the protoplanetary disk's evolution can become giant planets, which keep accreting gas as long as it is made available to them. Gas giant planets are mostly H and He by mass, although they can contain non-trivial amounts of heavier elements.

In classical formation models (Pollack et al. 1996), the growth of a giant planet undergoes three main phases: *Phase 1*, during which the accretion rate of solid material exceeds that of gas; *Phase 2*, during which the envelope mass increases at a rate larger than does the condensed core, until it becomes somewhat larger than the heavy-element core mass; *Phase 3*, during which rapid envelope contraction, and ensuing accretion of gas, leads to a (relatively) fast mass growth. These three phases are illustrated in Figure 2. Eventually, after the disk's gas in the planet's neighborhood disperses, the giant planet undergoes an *isolation phase*, during which it slowly contracts as it cools, possibly losing some of its gaseous component via evaporation driven by stellar irradiation if orbiting very close to the star.

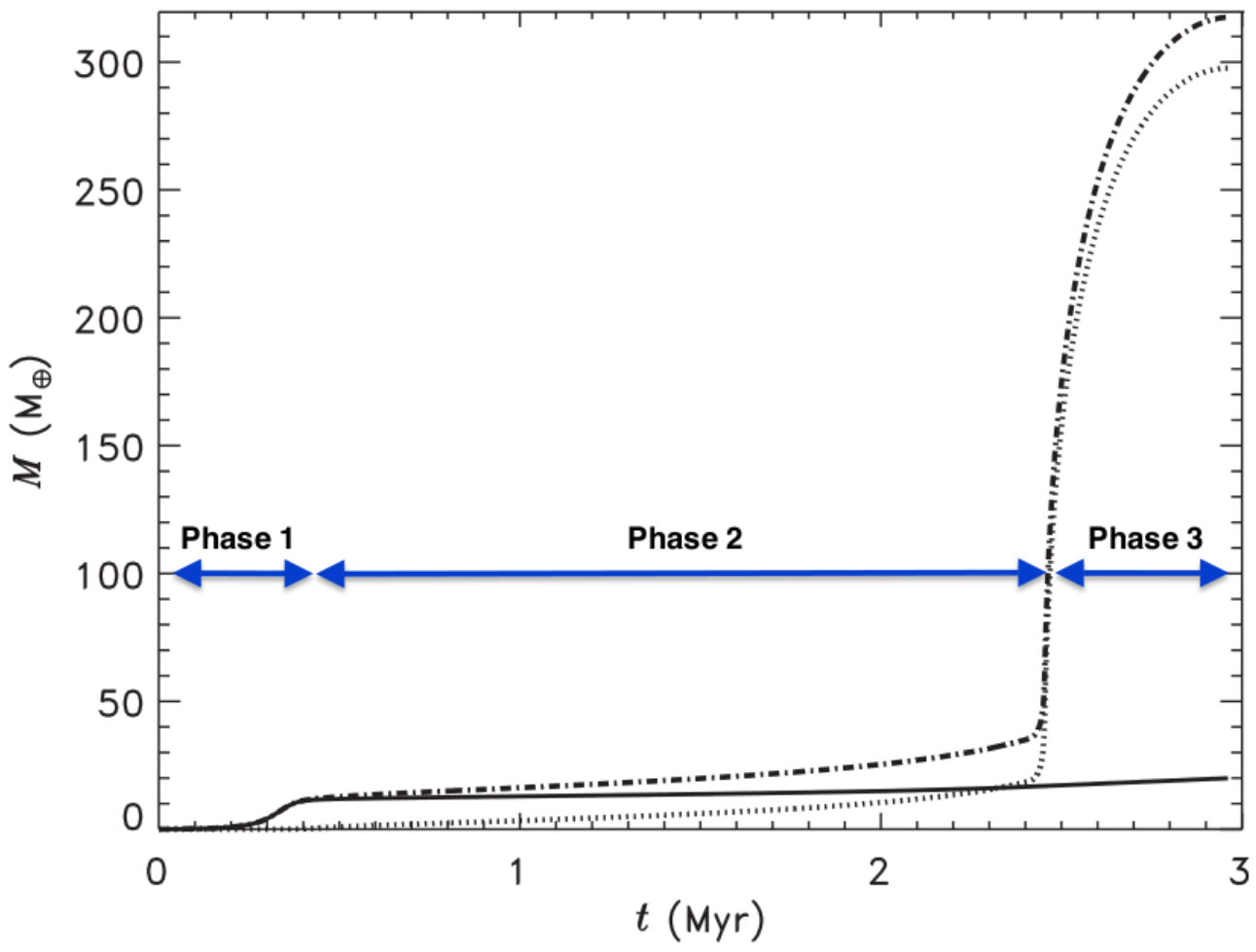

**Fig. 2** A formation model of Jupiter that shows the three main phases of the planet's growth prior to isolation (adapted from Lissauer et al. 2009). The solid line represents the mass of heavy elements in the condensed core. The dotted line represents the mass of H/He in the envelope. The dash-dot line indicates the total mass of the planet.

In this scenario, the main difference between a gas-rich planet and a gaseous giant planet is that the growth of the former is interrupted during Phase 1 or Phase 2, because of intervening gas dispersal or gas starvation by other means (e.g., by disk truncation). In the case of giant planets, gas starvation occurs during Phase 3. The time spent by a growing planet in Phase 2 depends on several factors, among which are the mass of the condensed core at the end of Phase 1 and the accretion rate of solids in Phase 2. The principal limitation of the classical model resides in the assumption that, during the accretion of solids, heavy elements release energy in the envelope but sink to the condensed core, neglecting the effects of their gas phase, which may have a non-trivial impact on the interior structure of both gas-rich and gas giant planets.

## *Core formation by accretion of planetesimals*

A collision between two planetesimals can lead to mass growth or mass loss, depending on the relative velocity of the impact, $v_{\mathrm{rel}}$. If growth is the typical outcome, the accretion rate is

$$\frac{dM}{dt} = \pi R^2 \Omega \Sigma_s F_g, \tag{1}$$

where $M$, $R$, and $\Omega$ are the mass, radius, and Keplerian angular velocity (about the star) of the growing body. For a swarm of planetesimals of surface density $\Sigma_s$ and thickness $H_s$, Equation (1) assumes that $|v_{\rm rel}| \sim H_s\Omega$, which applies when $H_s$ is much smaller than the local orbital radius. The product $\Sigma_s\Omega$ is the flux of solids through the cross section $\pi R^2$ and $F_g$ accounts for gravitational focussing, which acts to augment the geometrical cross section of the embryo. In the two-body-problem approximation (i.e., embryo and planetesimal), this enhancement factor is $F_g = 1 + (v_{\rm esc}/v_{\rm rel})^2$, where $v_{\rm esc} = \sqrt{2GM/R}$ is the escape velocity from the embryo. If $v_{\rm esc} \lesssim |v_{\rm rel}|$, $F_g \approx 1$ and the accretion rate in Equation (1) becomes proportional to the square of the planet's radius, $dM/dt \propto R^2 \propto M^{2/3}$. In contrast, if $|v_{\rm rel}| \ll v_{\rm esc}$, $F_g \propto R^2$ and $dM/dt \propto M^{4/3}$. However, in this low random velocity regime, three-body effects (which include the star) can limit the value of $F_g$ (Greenzweig and Lissauer 1992; Lissauer 1993). The growth of the body is drastically different in these two regimes. For $F_g \sim 1$, the accretion timescale $M/(dM/dt) \propto M^{1/3}$ increases with the mass, so small embryos double in mass faster than do large embryos and thus neighboring embryos tend to achieve similar masses. For $F_g \gg 1$, $M/(dM/dt) \propto M^{-1/3}$, so the largest embryo grows faster than the surrounding embryos, eventually accreting or scattering them, in a process referred to as and runaway growth (Wetherill and Stewart 1989).

Planetesimal-planetesimal encounters can excite orbital eccentricities and inclinations, but in a gaseous disk these are damped by gas drag on relatively short timescales (Adachi et al. 1976). Therefore, embryo and planetesimals can be assumed to travel on nearly circular orbits. In this case, $|v_{\rm rel}| \approx R_{\rm H}\Omega$, where $R_{\rm H}$ is the embryo's Hill radius. This represents the approximate distance inside which the embryo's gravity dominates that of the star and can be used as a measure of its gravitational reach (see also the review Chapter by P. Armitage). Since the gravitational enhancement factor reduces to $F_g \sim R_{\rm H}/R \gg 1$, the dominant embryo – the seed of the giant planet's core – can grow relatively quickly. This outcome is confirmed by direct calculations of an evolving swarm of planetesimals at 5.2 AU in a solar nebula, in which an embryo becomes a $3\,M_\oplus$ core in about $10^5$ years (e.g., Benvenuto et al. 2009; D'Angelo et al. 2014).

Planetesimals orbiting in a gaseous disk undergo drag-induced orbital migration. However, the timescale for orbital decay is much longer than that for eccentricity and inclination damping, and can exceed 1 Myr for 10–100 km size bodies, implying that the supply of planetesimals from larger distances is inefficient. Therefore, as a planetary core accretes bodies from a region around its orbit, know as feeding zone, this region gradually depletes and eventually empties. The half-width of the feeding zone is the maximum distance from the core's orbit at which planetesimals can be effectively deflected on core-crossing orbits. Calculations show that this distance is a few ($\approx 4$) times $R_{\rm H}$ (e.g., Mordasini et al. 2015). The mass acquired by a core prior to depletion of the feeding zone is

$$M_c \approx 4\pi a \Sigma_s (4R_{\rm H}) \approx \sqrt{\frac{(16\pi a^2 \Sigma_s)^3}{3M_\star}}, \tag{2}$$

where $a$ is the orbital radius of the core. This expression neglects the contribution of the envelope mass, which can enlarge $R_{\rm H}$. Provided that $\Sigma_s$ does not decline too quickly with orbital distance, $M_c$ is an increasing function of $a$ (Lissauer 1987). However, the growth timescale, $\propto 1/\Omega$, is also an increasing function of $a$.

A considerable depletion of the feeding zone marks the end of Phase 1. At this point, the envelope mass is typically small compared to the condensed core mass and the estimate of $M_c$ given in Equation (2) is valid. During Phase 2, as the planet's mass increases by accretion of gas, the feeding zone continues to expand, allowing additional heavy-elements to reach the planet. As the planet's mass grows larger and larger, however, effects such as scattering and capture in mean-motion resonances also affect the amount of solids available for accretion (e.g., Weidenschilling and Davis 1985). Radial migration of the planet can also alter this picture, since the feeding zone can extend into disk regions undepleted of solids (e.g., Alibert et al. 2004; Hasegawa and Pudritz 2012).

## *Core formation by accretion of small solids*

Planetary embryos may also grow via accretion of small, $\sim 1\,$cm–$1\,$m-size solids (Ormel and Klahr 2010; Lambrechts and Johansen 2012) if, at the epoch of formation, they represent a significant fraction of the total solids' surface density $\Sigma_s$. The actual size for efficient accretion depends on the embryo or planetary core mass and on the local thermodynamic conditions of the gas, hence it can vary with time and orbital distance. Gas in a protoplanetary disk is partially supported by its radial pressure gradient, so that its rotation rate at an orbital distance $a$ is $\approx \Omega\sqrt{1-(H_g/a)^2}$, where $H_g$ is the disk's pressure scale-height. If a solid particle orbits at the Keplerian rate $\Omega = \sqrt{GM_\star/a^3}$, its velocity relative to the gas is $\approx a\Omega(H_g/a)^2/2$. Although small (typically $\lesssim 10^{-3} a\Omega$), the resulting head-wind exerts an aerodynamic drag that removes orbital angular momentum from the particle, causing it to spiral toward the star on relatively short timescales, shorter than the timescale of the gas radial motion, $\sim a^2/\nu_g$, where $\nu_g$ is the kinematic viscosity of the gas (Pringle 1981).

The radial velocity of a particle is (e.g., Chiang and Youdin 2010)

$$\left|\frac{da}{dt}\right| \approx \frac{\Omega^2 (H_g/a)^2 \tau_s}{1+(\tau_s\Omega)^2} a, \tag{3}$$

in which $\tau_s$, the stopping time, depends on the drag coefficient, the density of both the gas and the solid, the particle size and relative velocity. The rate of change of $a$ in Equation (3) is maximum for $\Omega\tau_s \approx 1$, $|da/dt| \approx a\Omega(H_g/a)^2/2$. The accretion rate of solids through the orbital distance $a$ is then $dM_s/dt \approx 2\pi a \Sigma_s |da/dt|$. If $\Omega\tau_s \approx 1$, then $dM_s/dt \approx \pi H_g^2 \Sigma_s \Omega$, which corresponds to $\sim 5\times10^{-4}\Sigma_s/(1\,\mathrm{g\,cm^{-2}})(M_\oplus\,\mathrm{yr}^{-1})$ at 1 AU. An embryo growing at these rates would reach $\approx 10 M_\oplus$ in a few times $10^4$ years, or less. However, a growing planetary core accretes only at a fraction of $dM_s/dt$. Direct calculations indicate that the efficiency of accretion, i.e., the ratio of $dM/dt$ to $dM_s/dt$, depends on the core mass, particle size, and gas density (Kary et al. 1993). For a $1 M_\oplus$ core orbiting at around 1 AU, the accretion efficiencies of $\sim$ 1 cm–1 m particles range from a few to several percent (Morbidelli and Nesvorny 2012). Additionally, this heavy-element mass may not be readily transferred to the planet's interior because of rapid ablation of small solids very high up in the atmosphere (Alibert 2017). In fact, formation calculations suggest that this accretion route can directly form planetary cores of mass smaller, or much smaller, than $1 M_\oplus$ (depending on the composition of the accreted solids, Brouwers et al. 2018).

In this context, the feeding zone would be very extended. Because of the rapid radial drift of small solids, a core may continue to grow unless the supply of material was halted somehow. As density perturbations caused by tidal interactions between the growing core and the gas increase, drag forces can transfer angular momentum to particles exterior of the planet's orbit, opposing their inward motion. The strength of this effect depends on core mass, particle size, and gas temperature and density (e.g., Lambrechts et al. 2014).

## *Critical core mass and regimes of gas accretion*

Gas from the protoplanetary disk can become bound to a planetary core of mass $M_c$ if, in the frame of the core, its total energy becomes negative. Indicating with $u_{\mathrm{rel}}$ and $u_{\mathrm{th}}$, the relative (to the planet) and thermal velocities of the gas, this condition requires that $GM_c/R \geq (u_{\mathrm{rel}}^2 + u_{\mathrm{th}}^2)/2$. In this case, gas interior of the radius $R$ can become bound. Assuming that $u_{\mathrm{rel}}$ is dominated by Keplerian shear, this velocity is $|u_{\mathrm{rel}}| \approx R\Omega$ at distance $R$ from the core. The thermal velocity is $u_{\mathrm{th}} = c_g\sqrt{8/\pi}$, where $c_g$ is the sound speed of the gas. In a Keplerian disk, $c_g \approx H_g\Omega$. Therefore, as long as $R < H_g$, the inequality $u_{\mathrm{rel}}^2 \ll u_{\mathrm{th}}^2$ holds true and the radius

$$R_{\mathrm{B}} \approx a\left(\frac{M_c}{M_\star}\right)\left(\frac{a}{H_g}\right)^2, \tag{4}$$

known as the Bondi radius, is the maximum distance inside which gas can be bound to the core. Note that the condition $|u_{\mathrm{rel}}| < u_{\mathrm{th}}$ does not always apply, and the addition of the gas kinetic energy always results in a maximum distance for bound gas $< R_{\mathrm{B}}$. In particular, since the core's gravity perturbs the gas velocity in its proximity, the envelope radius is expected to be somewhat smaller than $R_{\mathrm{B}}$ (D'Angelo and Bodenheimer 2013). Examples of envelope formation around planetary cores are illustrated in Figure 3. The dashed lines indicate the size of the Bondi radius (see the figure's caption for further details). Three-body effects also limit the region of the bound gas, so that $R < R_{\mathrm{H}}$. Therefore, in general, the radius $R$ cannot exceed

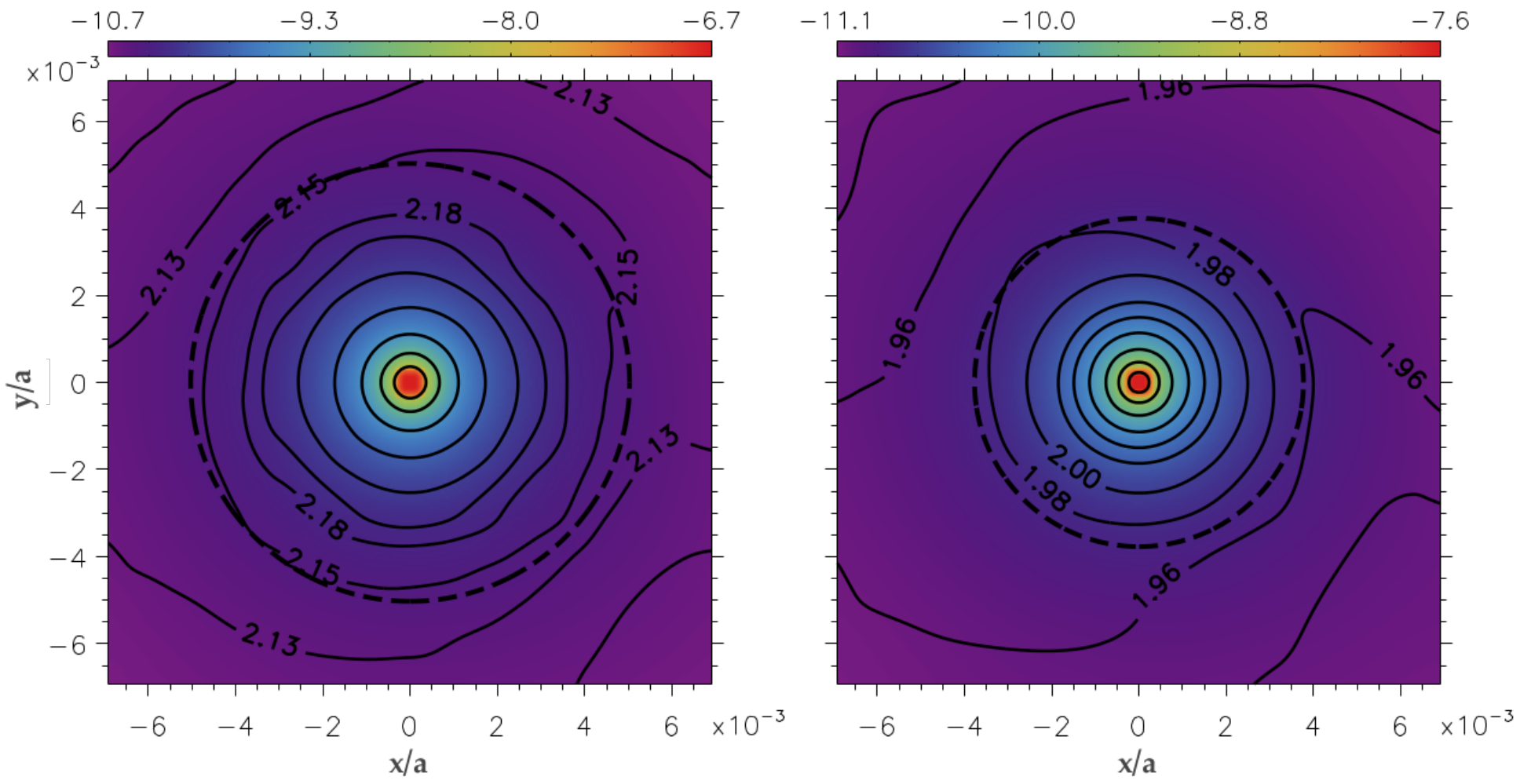

**Fig. 3** The envelopes surrounding a $5M_\oplus$ cores at 5 (left) and 10 AU (right) orbiting a solar-type star, obtained from three-dimensional, high-resolution radiation-hydrodynamical simulations (D'Angelo and Bodenheimer 2013). The color scale shows the gas density (on a logarithmic scale in units of $\mathrm{g\,cm^{-3}}$) in the disk's mid-plane. Temperature contours are also shown (on a logarithmic scale in units of K). For reference, the dashed circle has a radius equal to the the Bondi radius.

the smaller of $R_\mathrm{B}$ and $R_\mathrm{H}$ (in Figure 3, $R_\mathrm{H}/a \approx 0.017$). It should be stressed that the Bondi and Hill radii are two physically distinct and unrelated quantities, where the former is defined by an energy balance and the latter by a force balance.

As gas accretes on the core, and the envelope density and temperature increase, the resulting pressure gradient tends to oppose further accretion by inhibiting envelope contraction. Only by cooling, can the envelope accrete gas and grow. During the early stages of envelope accretion, energy is mainly provided by the accretion of solids, but work done by compression also contributes, becoming the dominant source of energy at later stages. During Phase 1 and most of Phase 2, the timescale for envelope growth is basically given by the Kelvin-Helmholtz timescale (e.g., Pollack et al. 1996; Ikoma et al. 2000; Ida and Lin 2004), which is the time that a gaseous envelope takes to lose its internal energy. If the planet is in near-hydrostatic equilibrium, this energy is related to the gravitational energy (Weiss et al. 2006; Kippenhahn et al. 2013) and the timescale can be approximated to $\tau_\mathrm{KH} \sim GM_cM_e/(R_eL)$, where $M_e$ and $R_e$ are the envelope mass and radius, respectively, and $L$ is the planet's luminosity. In envelopes whose outer shells are radiative, $L$ is regulated by the opacity of the medium, whose main contributor during these two phases is the entrained dust released by solids' accretion (e.g., Movshovitz et al. 2010).

The length of the cooling timescale during Phase 2, which may be comparable to the gaseous disk's lifetime ($\approx 2$ Myr in the model shown in Figure 2), represents an obstacle to the onset of rapid gas accretion and to the formation of a giant planet. However, it is consistent with Uranus and Neptune, and with the observed abun-dance of sub-Neptunes and gas-dwarf planets in general. In fact, these planets never underwent a proper phase of rapid gas accretion despite having masses of $10\,M_{\oplus}$, or more. In addition, the fact that Phase 3 has a somewhat limited duration, when it occurs, is also consistent with the rarity of very massive giant planets: there are substantially more planets in the mass range 1–$3\,M_{\rm J}$ than there are between 3 and $13\,M_{\rm J}$ (source: `http://exoplanet.eu`).

When contraction is slow, i.e., in the Kelvin-Helmholtz cooling limit, an envelope can be considered to evolve through quasi-hydrostatic equilibrium states, in which the pressure gradient (nearly completely) balances gravity. However, for a given core mass $M_c$, there is an envelope mass $M_e$ beyond which hydrostatic equilibrium can no longer exist, triggering the collapse of the envelope. Hence the concept of a critical core mass. There is a simple, yet elegant, theory to understand the physics of this process and to constrain the value of the critical core mass (Stevenson 1982). By constructing a hydrostatic envelope that is convectively stable, i.e., in which energy is transported only via radiation, under some simplifying assumption it can be shown that the largest core mass for which the envelope structure can remain hydrostatic is $M_c = 3M/4$, where $M = M_c + M_e$. Therefore, in this simplistic situation, once $M_e$ exceeds $M_c/3$, ensuing rapid contraction should initiate Phase 3.

Although the assumption of a fully radiative envelope may be accurate for tenuous atmospheres, it is not very representative of an envelope during Phase 2. Therefore, a simple theory for the critical core mass of a fully adiabatic, i.e., convective, envelope was also derived (Wuchterl 1993). Following similar arguments and applying the necessary approximations, it can be shown that the largest core mass for which the structure of a convective envelope remains hydrostatic is $M_c = 2M/3$. Therefore, in this case, the envelope mass must exceed $M_c/2$ to collapse and trigger rapid gas accretion. Although detailed calculations can result in multiple convective and radiative zones (e.g., D'Angelo et al. 2014), any working approximation of an envelope structure during Phase 2 should at least include an innermost convective zone and an outermost radiative layer (e.g., Piso et al. 2015). Using the above results, it follows that the largest core mass that allows for a hydrostatic structure of a convective-radiative envelope is

$$M_c = \left(\frac{2}{3}\right)\left(\frac{3}{4}\right)M = \frac{1}{2}M. \tag{5}$$

This is the often-quoted result that rapid envelope growth by gas accretion begins after the cross-over mass is achieved, that is when $M_e = M_c$ (e.g., Pollack et al. 1996; Hubickyj et al. 2005; Lissauer et al. 2009).

### *Effects of disk-planet tidal interactions*

During Phase 3 of giant planet formation, the rate of gas accretion can become so large that the protoplanetary disk may not be able to supply gas to the planet's vicinity at the required rate. Sufficiently far from a planet, the accretion rate through a steady-state viscous disk of mass $M_d$ is $\dot{M}_d \approx 3\pi\nu_g\Sigma_g$, where $\Sigma_g$ is the gas surface density. Closer to the planet, gravitational interactions may alter this result (Lubow and D'Angelo 2006). Therefore, the accretion of gas is typically limited by disk supply, especially in aged disks.

In order to better understand the process of disk-limited gas accretion, it is necessary to quantify the gravitational perturbations exerted by a planet on the surrounding gas. An evolving planet, say growing through Phase 2 or 3, exerts a gravitational torque on the gas given by $|\mathscr{T}_G| \sim \Sigma_g a^4 \Omega^2 (a/H_g)^3 (M/M_\star)^2$ (Goldreich and Tremaine 1980), whose radial distribution peaks at around $a \pm H_g$. The torque $\mathscr{T}_G$ is positive exterior and negative interior of the planet's orbit, with the net result of clearing gas from around the orbit. If this process is unimpeded, a density gap forms along the orbit. However, this process is opposed by the viscous torque, $\mathscr{T}_\nu \approx 3\pi\nu_g a^2 \Sigma_g \Omega$ (Pringle 1981), which acts to replenish gas and remove gaps. Consequently, when $\mathscr{T}_\nu \gg |\mathscr{T}_G|$ the disk remains smooth. In the opposite limit, that is when

$$\left(\frac{M}{M_\star}\right)^2 \left(\frac{a}{H_g}\right)^3 \gtrsim 3\pi \left(\frac{\nu_g}{a^2\Omega}\right), \qquad (6)$$

a gap can develop along the planet's orbit. The timescale for the formation of a gap is on the order of several tens of orbital periods of the planet. In typical protoplanetary disks around solar-type stars, a Saturn-mass planet at several AU can produce a shallow gap in the gas distribution. Around $\approx 0.1$ AU, even a Neptune-mass planet can produce a gap, due to the smaller value of the disk's aspect ratio $H_g/a$. In deriving the inequality in Equation (6), it is assumed that the torque associated to the pressure gradient is negligible compared to the other terms and that $H_g > R_\mathrm{H}$ when the gap starts to form. As the planet grows, the Hill radius eventually exceeds $H_g$ and the torque's radial distribution then peaks at around $a \pm R_\mathrm{H}$. Equation (6) reduces to $(M/M_\star) \gg \nu_g/(a^2\Omega)$ and the half-width of the deepest part of the gap becomes proportional to $R_\mathrm{H}$.

In early studies focussing on gap formation by a giant planets, it was suggested that the gap would separate the disk inside of the planet's orbit from the disk outside of it, effectively terminating gas accretion (Lin and Papaloizou 1986). Multi-dimensional hydrodynamic simulations later showed that, under typical disk conditions of temperature and viscosity, gas can still reach the planet and cross the gap (Kley 1999; Lubow et al. 1999; Kley et al. 2001). This behavior is shown in Figure 4. The lowermost and uppermost streamlines (e.g., Mihalas and Weibel Mihalas 1999) track the circulating motion of the gas, inside and outside the gap region. Gas streams that cross the gap (in either direction) and feed the planet's Roche lobe, are delimited by thicker curves. This gas approaches the planet, as indicated by the dark streamline, and is eventually accreted. The streamlines in this model are curves in three dimensions and the plot only shows their projections on the disk's mid-plane.

Although tidally-formed gaps are typically permeable to gas, their formation is very likely to impede gas accretion and thereby affect the final mass of a giant planet (Bryden et al. 1999; D'Angelo and Lubow 2008; Lissauer et al. 2009). In fact, for given disk conditions, the disk-limited gas accretion rate of a planet in Phase 3 increases with the mass $M$ until it attains a maximum, when right-hand and left-hand sides of Equation (6) are approximately equal (Bodenheimer et al. 2013), and then it declines as $M$ increases further.

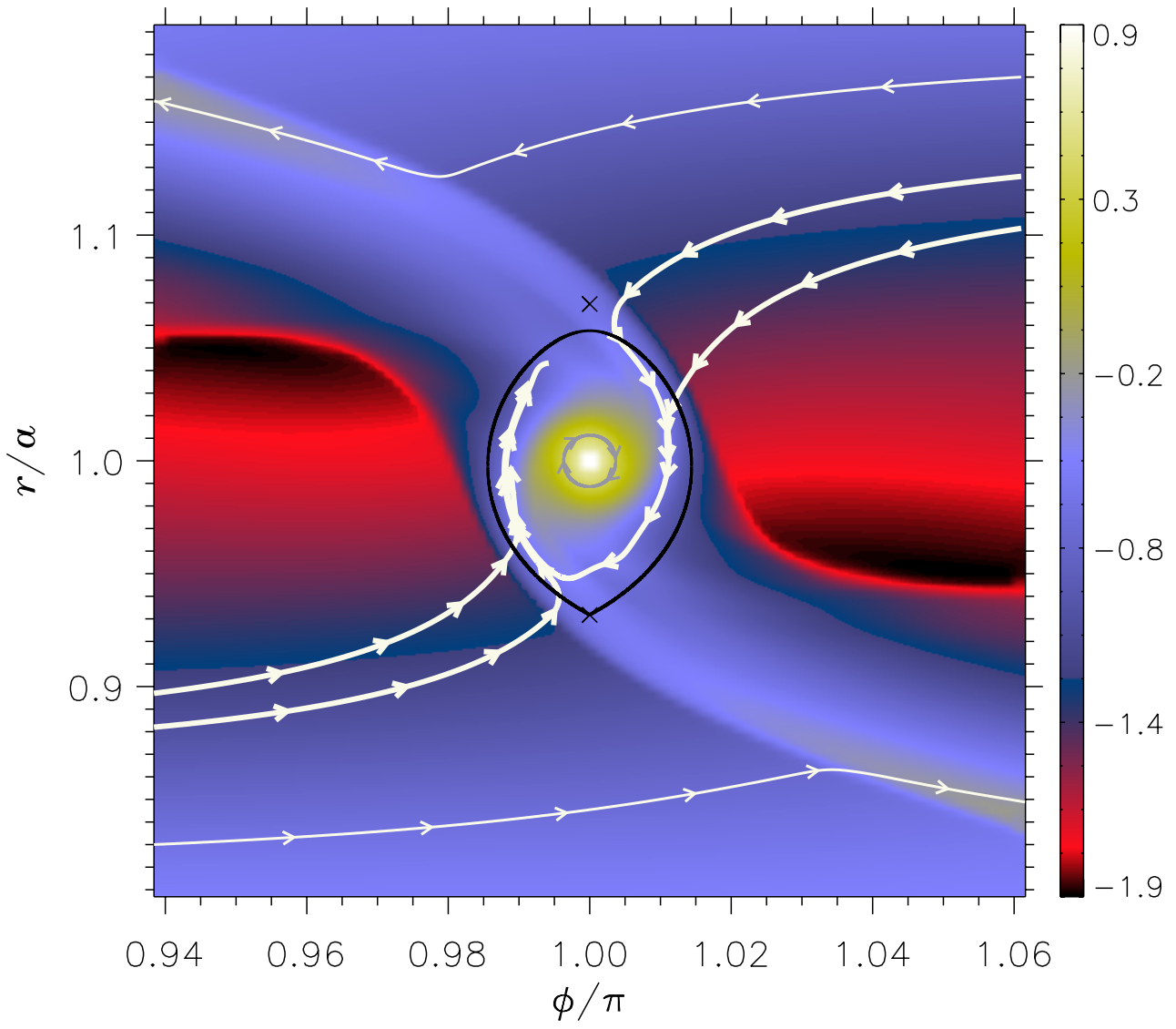

**Fig. 4** The color scale represents the logarithm of the gas surface density in a disk in the vicinity of a Jupiter-mass planet, calculated from a three-dimensional model (D'Angelo and Podolak 2015). The surface density is normalized to that of the unperturbed disk at the planet's position, $(r = a, \phi = \pi)$. The white curves are gas streamlines, in the frame co-rotating with planet, projected on the disk's mid-plane. The black curve is the trace of the planet's Roche lobe (the Lagrange points $L_1$ and $L_2$ are also plotted, e.g., Murray and Dermott 1999). See text for further discussion.

As gas streams through a deep gap toward the planet, it can form a circumplanetary disk, that is a disk orbiting the giant planet. These disks also carry solids and are likely the formation sites of satellites. They are thought to have existed around Jupiter and Saturn toward the end of their accretion history, as suggested by the orbital properties of their natural satellites (e.g., Coradini et al. 1981). Compositional gradients of Jupiter's natural satellites also contribute some evidence (e.g., Peale 2007). An example of the formation of a circumplanetary disk around a Jupiter-mass planet is displayed in Figure 5. The figure shows the gas density in a vertical section of the disk, which occupies the inner portion of the planet's Roche lobe. In case of Jupiter, $0.1\,R_{\rm H}$ is equal to $\approx 75$ Jupiter's radii.

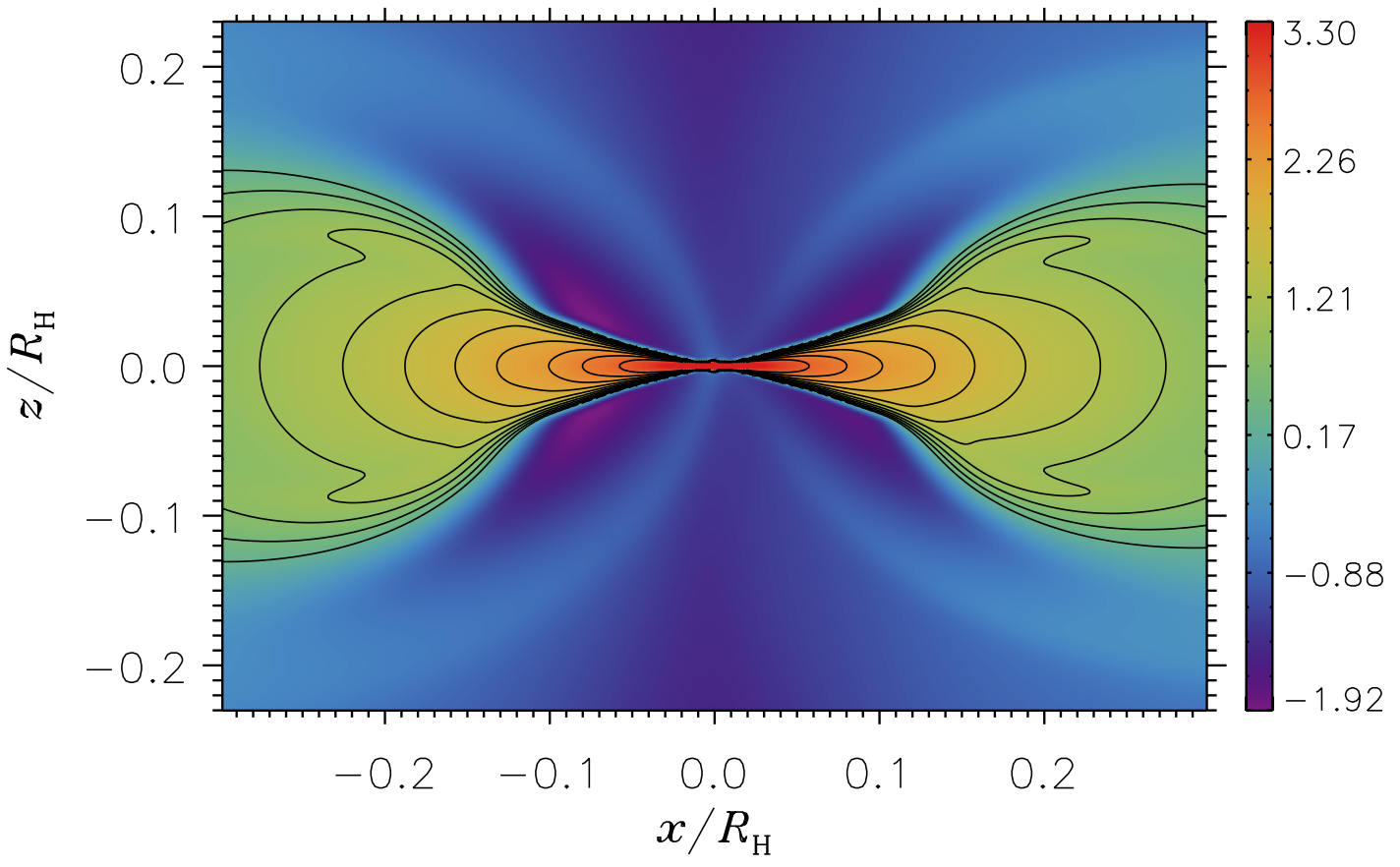

**Fig. 5** The circumplanetary disk around a Jupiter-mass planet orbiting in a protoplanetary disk, obtained from a three-dimensional, high-resolution hydrodynamic simulation (D'Angelo and Podolak 2015). The color scale shows the logarithm of the gas density (and some density contours) in a vertical slice passing through the planet. The density is normalized to that of the unperturbed disk at the location of the planet (e.g., $10^{-11}\,\mathrm{g\,cm^{-3}}$ for an unperturbed surface density of $100\,\mathrm{g\,cm^{-2}}$).

## *Disk removal, final mass, and evolution in isolation*

As mentioned above, the gas component of a protoplanetary disk lasts a few to several million years and the phase of slow contraction of a giant planet's evolution (Phase 2) can be of comparable duration. It is thus natural to conjecture that, once entered Phase 3, the mass growth of a giant planet is first limited by gap formation and is then terminated by intervening dispersal of gas around the planet's orbit. In this scenario, Saturn is particularly revealing, since its mass may place it around the maximum of the disk-limited accretion curve (Lissauer et al. 2009). Hence, Saturn's mass requires gas dispersal, which also implies that the condensed cores of Uranus and Neptune simply took too long to form and the planets did not have time to evolve far along Phase 2.

Jets and winds can remove gas from young protoplanetary disks. But as a disk settles in a more quiescent state, viscous transport toward the star becomes the main driver of its evolution. During the planet formation epoch, and until gas is eventually dispersed, viscous diffusion and gas removal by irradiation from the central star play the major role in determining the disk's lifetime. X-ray, extreme-ultraviolet (EUV), and far-ultraviolet (FUV) radiation can heat the disk's surface layers and unbind the gas, generating a thermal wind, a process known as disk photo-evaporation. Mass loss interior of a few AU is typically determined by EUV radiation, and by FUV and X-ray photons at greater distances. Global time-averaged photo-evaporation rates are in the range between $\sim 10^{-9}$ and $\sim 10^{-8}\,M_{\odot}\,\mathrm{yr}^{-1}$ (Gorti et al. 2016).

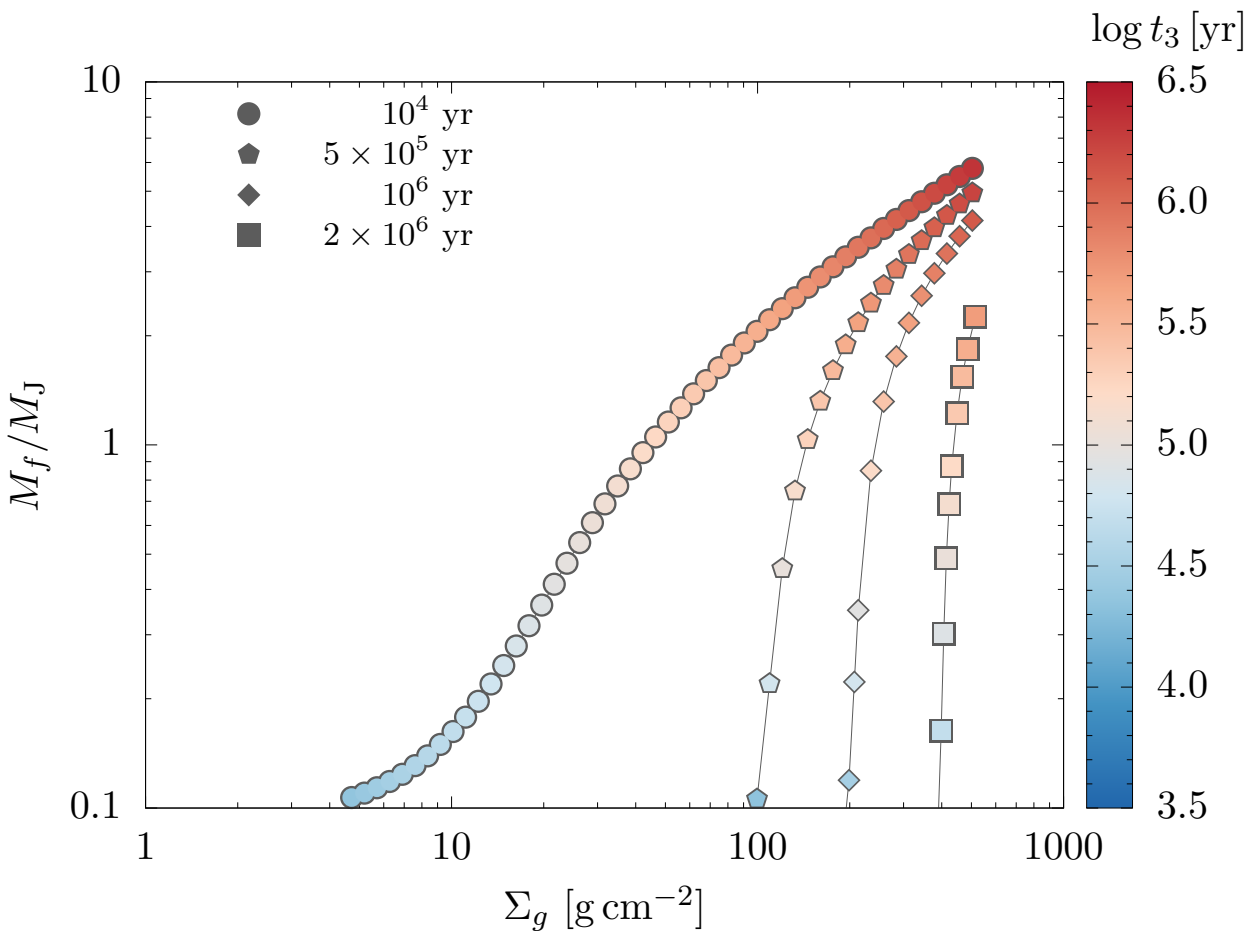

**Fig. 6** Examples of giant planets' masses after formation versus the gas surface density around the planet's location at the end of Phase 1. Gas disperses via viscous diffusion and photo-evaporation by stellar irradiation. Disk-limited accretion rates are obtained through three-dimensional hydrodynamic calculations of disk-planet interactions (Bodenheimer et al. 2013). The color scale indicates the duration of Phase 3, $t_3$, that is the time between the beginning of Phase 3 and gas dispersal. Different symbols mark different lengths of Phase 2, as indicated in the legend. See text for further details.

Disk evolution by viscous diffusion and dispersal by photo-evaporation can be combined with numerically determined disk-limited gas accretion rates (e.g., Bodenheimer et al. 2013) to estimate the final masses of giant planets, $M_f$ (see, e.g., Hellary and Nelson 2012; Hasegawa and Pudritz 2013; Dittkrist et al. 2014, for a more general approach). A solution for the viscous evolution of a disk of mass $M_d$ yields a steady-state accretion rate (Lynden-Bell and Pringle 1974)

$$\dot{M}_d \approx \frac{\dot{M}_d^0}{(4\dot{M}_d^0 t/M_d^0 + 1)^{5/4}}, \tag{7}$$

where $M_d^0$ and $\dot{M}_d^0$ are respectively the disk mass and accretion rate at a reference time, here roughly corresponding to the end of Phase 1. The gas removed by photo-evaporation (and accretion on the planet) can be added to that removed by viscous diffusion to obtain the disk mass $M_d$ at a given time. The surface density of the gas is estimated by assuming a power-law decline with orbital radius. During Phase 2, the planet is assumed to accrete gas until $M_e > M_c$, at which point Phase 3 begins.

Figure 6 shows examples of the dependence of the final planet mass, $M_f$, on the value of $\Sigma_g$, the local gas density at the end of Phase 1, for various lengths of Phase 2 (see legend). The tracks in the figure assume that the planet orbits a solar-mass star. The color scale indicates the length of Phase 3, $t_3$,which ends when gas disperses from around the planet's orbit. The upper-most curve represents a case in which the duration of Phase 2 is very short, as it may be the case when solids' accretion stops abruptly, ceasing Phase 1 and readily allowing sustained gas accretion. This might occur, for example, if core accretion is dominated by small solids. The track represented by squares would predict the formation of a Jupiter-mass planet when the gas surface density at the beginning of Phase 3 is $\approx 10\,\mathrm{g\,cm^{-2}}$ and $t_3 \approx 2 \times 10^5$ years, and of a Saturn-mass planet when the gas density is $\approx 5\,\mathrm{g\,cm^{-2}}$ and $t_3 \approx 10^5$ years. The results in the figure use a turbulence parameter $\alpha_g$ (Shakura and Sunyaev 1973) of a few times $10^{-3}$. Since the kinematic viscosity of the gas affects disk-limited gas accretion, $M_f$ can be different for significantly different values of $\alpha_g$ (Lissauer et al. 2009; Bodenheimer et al. 2013).

Once gas around the planet dissipates (via photo-evaporation or some other process), the planet evolves in isolation at a constant mass. The planet gradually cools and contracts on the Kelvin-Helmholtz timescale. The thermal state of the envelope at the end of the formation epoch, determined by the accretion history during Phase 3, also influences part or most of the post-formation evolution. For planets whose mass is $\lesssim 10\,M_\mathrm{J}$, the luminosity is provided by the release of gravitational energy during compression. For more massive planets, nuclear burning of deuterium represents a significant source of energy as well. Planets orbiting very close to their star may lose gas by absorption of high-energy photons emitted by the star (e.g., Ehrenreich and Désert 2011; Salz et al. 2016). Theories based on energy-limited hydrodynamical escape predict that this loss is proportional to $\pi R^2 \mathscr{F}_\star [R/(GM)]$, where $\mathscr{F}_\star$ is the incident flux of X-ray and EUV photons. Gas loss is then proportional to $R^3$ and inversely proportional to $a^2$ ($\mathscr{F}_\star \propto a^{-2}$) and, therefore, it is most effective within the first few hundred million years of the planet's evolution in isolation. Additionally, late impacts may also alter somewhat the heavy-element content of giant planets (Ginzburg et al. 2017).

## Formation by Disk Instability

Core formation at large distances from a star requires timescales that may be too long to initiate Phase 3. Yet, a number of massive giant planets have been directly imaged far from their stars, as is the case of the four-planet system orbiting the young star HR 8799, displayed in Figure 7 (e.g., Bowler 2016).

Historically, even before the introduction of the core nucleated accretion scenario, the formation of a giant planet was thought to be akin to that of a star: a gravitational instability occurring in the gaseous medium would lead to fragmentation and to the formation of a self-gravitating clump, a newly born gaseous planet (Kuiper 1951). In this case, the formation would be rapid, taking a few to several orbital timescales, although the planet would initially be very large and contraction to roughly jovian size would take far longer (Cameron et al. 1982). Since the gaseous medium is a protostellar disk, this mechanism is referred to as formation by disk instability.

Revived two decades ago (Boss 1997), this mode of formation has been extensively studied afterward (e.g., Durisen et al. 2007; Durisen 2011), and the prevailing consensus is now that it probably favors the formation of massive objects, brown dwarfs or low-mass stellar companions (e.g., Kratter and Lodato 2016). However, if indeed disk instabilities induced fragmentation, leading to giant planet formation, they would likely do so in the outskirts of protoplanetary disks (e.g., Durisen 2011), offering a possible explanation for the imaged planetary-mass objects. A substantial effort is currently underway to test the viability of the disk instability as a possible formation scenario for gaseous planets (e.g., Helled et al. 2014).

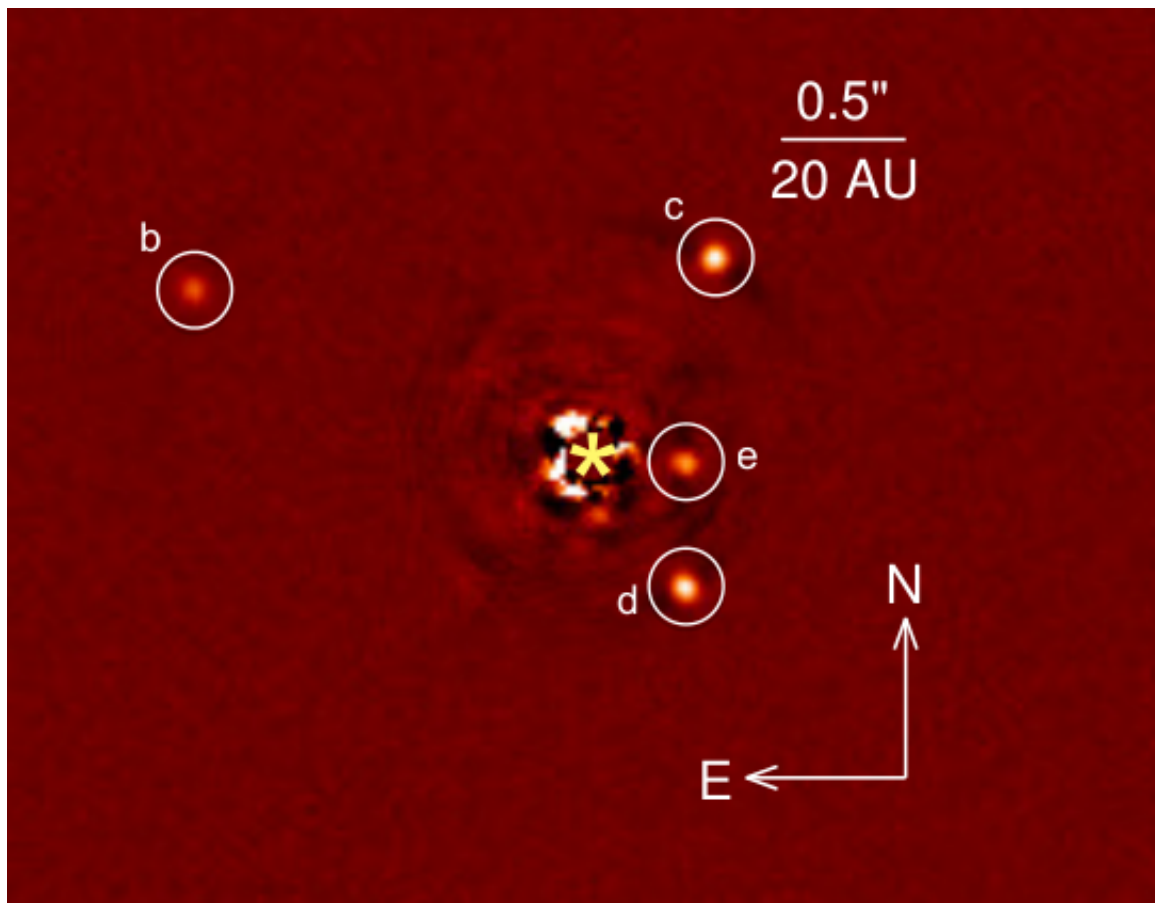

**Fig. 7** The image shows four massive planets ($M = 5$–$10\,M_J$) harbored by the star HR 8799. The planets' orbital distances are between 15 and 70 AU (Marois et al. 2008, 2010). The image was taken in $L'$ band ($3.8\,\mu$m) by the LMIRCam camera mounted on the Large Binocular Telescope (adapted from Maire et al. 2015).

## *Gravitational instabilities and disk fragmentation*

Gravitational instabilities arise in a differentially rotating disk when the destabilizing effect of gas self-gravity overcomes, locally, the stabilizing effect of gas rotation and pressure (Durisen 2011). In a thin Keplerian disk, a condition for instability is expressed by the inequality

$$\mathscr{Q} = \frac{c_g \Omega}{\pi G \Sigma_g} < \mathscr{Q}_{\rm crit}, \tag{8}$$

in which $\mathscr{Q}$ is known as the Toomre stability parameter (Toomre 1964, see also Safronov 1960) and $\mathscr{Q}_{\rm crit}$ is a critical value, equal to 1 in case of instabilities driven by axisymmetric (i.e., ring-like) perturbations (Binney and Tremaine 1987). More generally, for non-axisymmetric (e.g., spiral-like) perturbations, $\mathscr{Q}_{\rm crit}$ is somewhat larger, but $\lesssim 2$. Since $c_g^2 \propto T_g$, the gas temperature, and $\Sigma_g$ is proportional to the local disk mass, the condition (8) implies that cold and massive regions of a disk are more susceptible to becoming gravitationally unstable ($\mathscr{Q} \propto \Omega \sqrt{T_g}/\Sigma_g$). Once the inequality (8) is satisfied, a gravitational instability can grow on a timescale $2\pi/\Omega$.

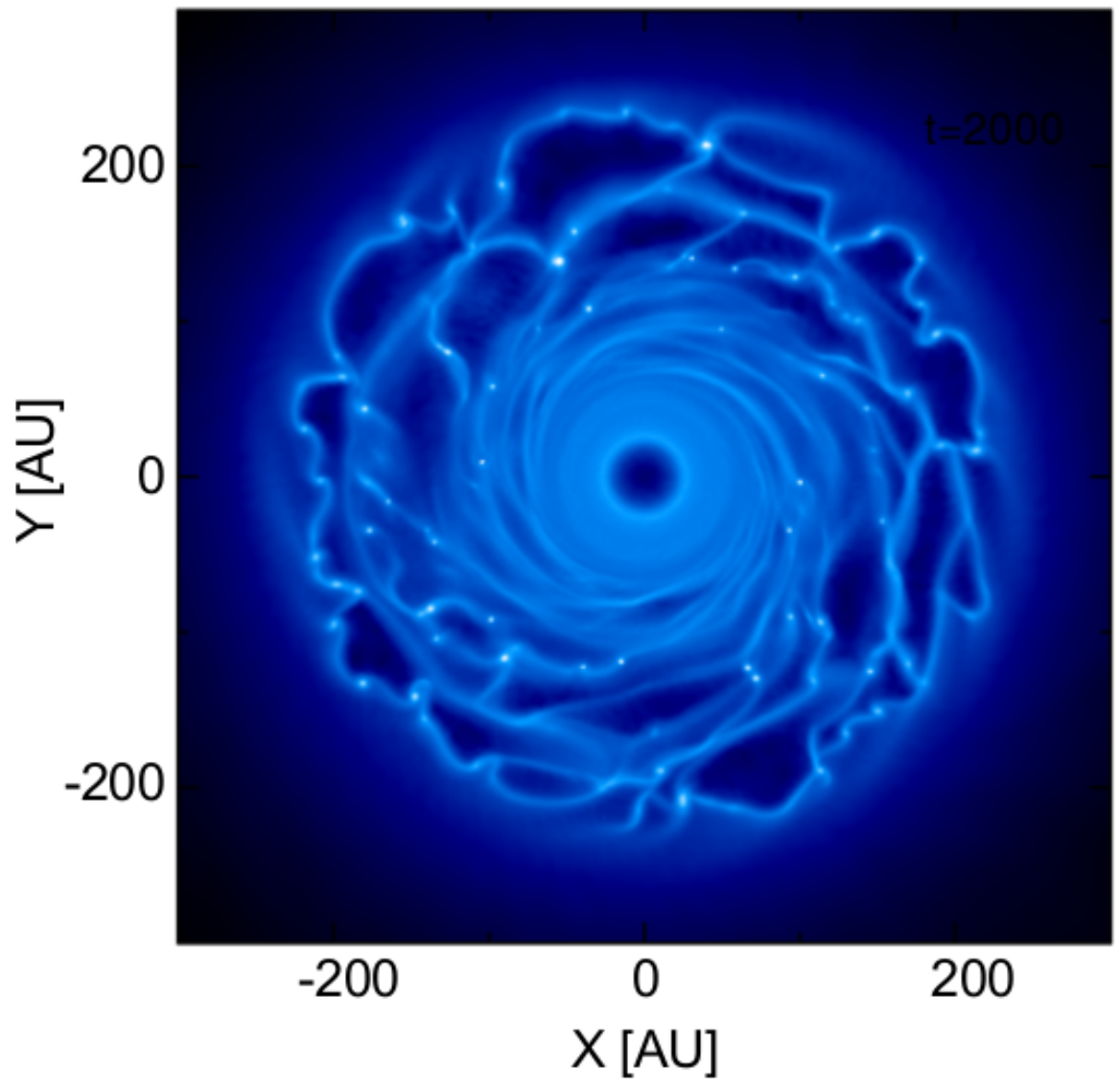

**Fig. 8** An example of hydrodynamical calculation of a protostellar disk around a $1 M_{\odot}$ star that undergoes gravitational instability resulting in fragmentation (adapted from Forgan et al. 2017). The color scale renders the mean column density. The disk mass is $0.25 M_{\odot}$ and $\Sigma_g$ drops as the inverse of the distance from the star. The cooling timescale of the gas is $4/\Omega$. In this case, clumps tend to survive several orbital periods before being tidally disrupted.

The evolution of gravitational instabilities in disks is controlled by gas cooling. If the local cooling timescale is somewhat larger than $2\pi/\Omega$, the internal energy produced by the instability and heat loss nearly balance each other out. In this case, the instability may be sustained in a nearly steady-state (Gammie 2001). However, if the cooling timescale becomes $\lesssim 2\pi/\Omega$, the balance is broken and the disk can fragment into self-gravitating clumps. Numerical studies have established that fragmentation sensitively depends on the equation of state of the gas and on its cooling history (e.g., Rice et al. 2003), but also on the numerical treatment of the system, e.g., on the adopted geometry (two versus three dimensions, Young and Clarke 2015) and on the computation of the energy transfer within the gas (Rogers and Wadsley 2011). Additionally, due to inefficient cooling, both analytic arguments and numerical simulations indicate that the possibilities of fragmenting a disk inside several tens of AU appear remote under reasonable circumstances (e.g., Rafikov 2005; Rogers and Wadsley 2011; Durisen 2011).

The radial extent of a disk region subject to gravitational instability is $\sim 2\pi H_g/\mathscr{Q}$. If fragmentation ensues, several clumps are spawned, each of which may have a

mass

$$M \sim \pi a^2 \Sigma_g \left(\frac{H_g}{a}\right). \tag{9}$$

In a massive disk, this mass may equal a few to tens of Jupiter's masses at hundreds of AU, although interactions among clumps may quickly alter these values. The initial clump size would be enormous, $\approx R_{\rm H}$, or $\approx 7\,$AU if a $1\,M_{\rm J}$ clump formed at 100 AU. An example of a fragmenting disk is illustrated in Figure 8. In the case shown, clumps form beyond $\approx 50\,$AU and have masses between $\approx 1$ and $\approx 4\,M_{\rm J}$, although they tend to be short-lived. The survival of a clump is tied to its ability to cool and contract before it is disrupted by interactions with other clumps or other density perturbations, e.g., other fragments, or it is tidally shredded by the gravity of the star if it drifts inward too quickly. Unfortunately, simulating the evolution of these clumps is quite challenging, because of resolution limits in grid-based calculations and numerical noise in particle-based calculations (Young and Clarke 2015).

## *Heavy-element enrichment*

A giant planet born out of a clump produced by gravitational instability has an initial composition equal to that of the disk's gas, i.e., its metallicity is $Z \approx 0.01$ (see also the review Chapter by T. Bergin). Dust and particles contained in the gas would sediment toward the center of the collapsing clump on a timescale shorter or somewhat shorter (depending on the grain size and on internal gas turbulence) than the contraction timescale, possibly producing a condensed core of a few Earth's masses, if $M \approx 1\,M_{\rm J}$. However, the metallicity of the clump may be super-stellar to begin with, since spiral density perturbations can sweep up dust, through aerodynamic drag, and concentrate it along their density maxima prior to fragmenting. In this case, the increased metallicity of the clump may also affect its contraction and survival (e.g., Kratter and Lodato 2016).

Jupiter and Saturn are substantially enriched in heavy elements compared to the Sun (e.g., Wong et al. 2004; Hersant et al. 2008), and some observed extrasolar giant planets too appear to have super-stellar metallicities (e.g., Sato et al. 2005; Miller and Fortney 2011; Jordán et al. 2014). An enhancement in the metal content may also occur in planets formed through disk fragmentation. In fact, planetesimals can be accreted in a similar manner as done by planetary cores and described previously. If fragmentation produced a clump of mass $M \approx 1\,M_{\rm J}$ at 50 AU, where $\Sigma_g \approx 10\,{\rm g\,cm^{-2}}$, then a surface density of solids $\Sigma_s \approx 0.1\,{\rm g\,cm^{-2}}$ might endow the planet with additional $\approx 10\,M_\oplus$ of heavy elements (see Equation (2)). These estimates assume that the planet empties its feeding zone and that the accretion occurs when the disk has settled in a more quiescent state (i.e., well after clump formation). In fact, in a gravitationally unstable disk, density fluctuations can excite the orbital eccentricity of planetesimals, increasing their velocity relative to the planet and hindering accretion. Moreover, direct calculations indicate that there is significant scattering of planetesimals, toward the inner and outer disk, reducing the amount available for accretion (Walmswell et al. 2013).

Whether originally present in the clump or accreted afterward, solids tend to settle, possibly forming a condensed core. In this case, the core formation is a consequence of the planet's evolution, but it is completely unnecessary for the formation of a giant planet. This is perhaps the biggest difference with the core nucleated accretion scenario, which requires a condensed core *to form* a giant planet, even though, after giga-years of evolution, its presence may be difficult to ascertain or even define!

## *Migration and downsizing*

Planets formed from clumps will be subjected to strong gravitational torques by the massive disk in which they are generated. These torques will change the planet's orbital angular momentum, leading to orbital migration. Consider a gravitationally unstable disk, fragmenting at $a \approx 100\,\mathrm{AU}$ and forming a $M \approx 3\,M_{\mathrm{J}}$ planet ($\Sigma_g \approx 10\,\mathrm{g\,cm^{-2}}$). For a gas viscosity corresponding to a parameter $\alpha_g \approx 0.1$, a typical value measured in simulations of strongly self-gravitating disks, and $H_g/a \approx 0.1$, the kinematic viscosity would be $\nu_g = \alpha_g H_g^2 \Omega \approx 10^{-3} a^2 \Omega$. Therefore, the criterion (6) for gap formation would not be satisfied and the planet would undergo a regime of migration referred to as *Type I*, in which the applied torque is proportional to the local gas density and to the square of the planet's mass (see the review Chapter by R. Nelson). In a smooth disk, the rate of change of the orbital angular momentum would lead to inward migration at a rate

$$\left|\frac{da}{dt}\right| \sim \left(\frac{a^2 \Sigma_g}{M_\star}\right) \left(\frac{M}{M_\star}\right) \left(\frac{a}{H_g}\right)^2 a\Omega. \tag{10}$$

For the values stated above, $|da/dt| \approx 0.003 a\Omega$ and the orbital decay time would be tens of orbital periods. Direct simulations confirm that migration is very fast and generally toward the star (e.g., Baruteau et al. 2011), possibly stalling once the planet reaches the threshold mass for deuterium burning ($\approx$ 12–14 $M_{\mathrm{J}}$), or beyond (Stamatellos 2015).

Two consequences arise from the rapid orbital decay of giant planets formed in gravitationally unstable disks, if they survive. The first is that, even though they tend to form far away from their stars, they can rapidly move to much shorter distances, offering a possible interpretation of the giant planets observed at orbital distances of several AU or less. The second is tidal stripping and downsizing. Prior to significant contraction, the radius of a planet is equal to (some fraction of) its Hill radius, i.e., $R \propto R_{\mathrm{H}} = a[M/(3M_\star)]^{1/3}$ ($< R_{\mathrm{B}}$ in Equation (4)), and cannot exceed it. Henceforth, over a timescale shorter than the contraction timescale, the radius evolves so that $dR/R = da/a + dM/(3M)$. Therefore, for $dM \lesssim 0$, the planet shrinks as it migrates toward the star. Physically, the planet's outer layers end up around or outside the boundary of the planet's Roche lobe, are subject to the now dominating stellar gravity, and become unbound. The dynamics of the disk's gas around the planet can also contribute to unbind gas. This process may reduce the mass of planets formed by gravitational instability and has been invoked as a possible explanation of small-mass gaseous, and even mostly rocky, planets observed in close proximity of stars. If a planet with a core moves very close to the star, say within a few times 0.01 AU, having shed most of its envelope along the way, it may lose the remainder of its atmosphere by stellar irradiation and leave a bare condensed core (Helled et al. 2014). However, by the same mechanism, a clump or planet may simply be disrupted, which is often the case (Forgan and Rice 2013). This argument applies when the contraction timescale is relatively long. But if it is comparable to the migration timescale, a migrating planet may actually gain mass during contraction, in a similar fashion as do planets formed by core nucleated accretion during Phase 3. This outcome has actually been observed in hydrodynamical simulations (e.g., Stamatellos 2015).

## Challenges to Formation Models

Any giant planet formation theory must be applicable to the solar system giants. Although detailed measurements of Jupiter's interior are currently being delivered by the Juno spacecraft, uncertainties remain large. Likewise, data taken during Cassini's Grand Finale, the final orbits of the Cassini spacecraft, may provide better estimates of Saturn's gravity field and internal structure. The core nucleated accretion model offers a reasonably good description – according to the current knowledge – of observed giant planets, although it may face difficulties to account for the existing population of extrasolar planets on very wide orbits. For these somewhat uncommon planets, formation via gravitational instability may offer a viable alternative, and the possibility of diverse modes of formation may be suggested by observations (Santos et al. 2017). Nonetheless, to date, the prospects that close-in giant (or smaller) planets may result from clumps formed in gravitationally unstable disks appear unfavorable (Forgan and Rice 2013).

The core nucleated accretion model has progressed substantially over the past two decades, by accounting for an ever growing number of physical effects and becoming a mature theory. It is now regarded as the principal mode of giant planet formation. Likewise, a substantial effort has been made to improve the disk instability model over recent years. Nonetheless, both scenarios necessitate refinement and further development. For example, the former needs to address the importance and consequences of compositional gradients and internal mixing, whereas the latter needs to better pin down the physical conditions conductive of fragmentation and clump survival. The interior structures of planetary-mass objects born out of gravitational collapse, including their inventory of heavy elements, are largely unknown and will need to be studied and constrained in the future.

Clearly, in either case, future improvements must be dictated and guided by well-established observational results. It is expected that in the near future, as the quantity and quality of data keeps growing, both scenarios will face increasingly demanding observational tests. At this stage, it is important that models focus on what data indicate to be common outcomes of the evolution process.

## Cross-References

- Extrasolar Planets Population Synthesis (Mordasini, C.)
- Connecting Planetary Composition to Formation (Pudritz, R., Cridland, A., Alessi, M.)
- A Brief Overview of Planet Formation (Armitage, P.)
- Chemistry During the Gas-Rich Stage of Planet Formation (Bergin, T.)
- Planetary Migration in Protoplanetary Disks (Nelson, R.)

**Acknowledgements** This work benefitted greatly from discussions with Peter Bodenheimer. The authors acknowledge support from NASA's Research Opportunities in Space and Earth Science (ROSES), and in particular from the Emerging Works Program. Resources supporting the work shown in Figures 1 through 6 were provided by the NASA High-End Computing (HEC) Program through the NASA Advanced Supercomputing (NAS) Division at Ames Research Center.